# Successful Combination of Database Search and Snowballing for Identification of Primary Studies in Systematic Literature Studies

Claes Wohlin *,a, Marcos Kalinowski b, Katia Romero Felizardo c, EmiliaMendesa

aBlekinge Institute of Technology, Karlskrona, Sweden
bPontifical Catholic University of Rio de Janeiro, Rio de Janeiro, Brazil
cFederal Technological University of Paraná, Cornélio Procópio, Brazil

**Abstract**

**Background**: A good search strategy is essential for a successful systematic literature study. Historically, database searches have been the norm, which has later been complemented with snowball searches. Our conjecture is that we can perform even better searches if combining the two search approaches, referred to as a hybrid search strategy.
**Objective**: Our main objective was to compare and evaluate a hybrid search strategy. Furthermore, we compared some alternative hybrid search strategies to assess whether it was possible to identify more cost-efficient ways of searching for relevant primary studies.
**Method**: To compare and evaluate the hybrid search strategy, we replicated an SLR on industry–academia collaboration in software engineering. The SLR used a more "traditional" approach to searching for relevant articles for an SLR, while the replication was conducted using a hybrid search strategy.
**Results**: In our evaluation, the hybrid search strategy was superior in identifying relevant primary studies. It identified 30% more primary studies and even more when focusing only on peer-reviewed articles. To embrace individual viewpoints when assessing research articles and minimise the risk of

*Corresponding author
  Email addresses: `claes.wohlin@bth.se` (Claes Wohlin *),
`kalinowski@inf.puc-rio.br` (Marcos Kalinowski), `katiascannavino@utfpr.edu.br` (Katia Romero Felizardo), `emilia.mendes@bth.se` (Emilia Mendes) missing primary studies, we introduced two new concepts, *wild cards* and *borderline articles*, when conducting systematic literature studies.

**Conclusions**: The hybrid search strategy is a strong contender for being used when conducting systematic literature studies. Furthermore, alternative hybrid search strategies may be viable if selected wisely in relation to the start set for snowballing. Finally, the two new concepts were judged as essential to cater for different individual judgements and to minimise the risk of excluding primary studies that ought to be included.

## 1. Introduction

According to the guidelines for performing systematic literature studies authored by Kitchenham and Charters (2007), a secondary study is defined as "A study that reviews all the primary studies relating to a specific research question with the aim of integrating/synthesising evidence related to a specific research question."

Given that the definition highlights that we are expected to find "all" the primary studies, the search strategy becomes essential in achieving (or at least move towards) this goal. The two main approaches to search for primary studies are database search and snowballing. The Kitchenham and Charters (2007) guidelines describe database search, and Wohlin (2014) provides guidelines for using snowballing. These two search strategies are compared by Jalali and Wohlin (2012). However, an alternative is to combine them. The concept of a hybrid approach was proposed by Wohlin (2014) and listed as an area for future research. Since then, Mourão et al. (2017) investigated one hybrid search strategy. This was then extended to four alternative hybrid search strategies by Mourão et al. (2020), and they were also evaluated.

A hybrid search strategy is *defined* herein as follows: A hybrid search strategy is the pre-planned integration of at least two systematic approaches to searching for articles for a systematic literature review (SLR). For example, conducting systematic searches in one or more digital databases or indexing services and then snowballing from all the relevant articles found in the previously conducted searches. A hybrid search strategy should not be mixed up with conducting several complementary searches, for example, a database search complemented with searching in some specific conference proceedings. Furthermore, to conduct a database search and then to conduct a snowballing from a sample of the articles found in the database search is also not a hybrid search strategy according to our definition.

At the outset of the research, the overall objective was to compare and evaluate an SLR using one search strategy with a replication using a hybrid search strategy and compare the four different hybrid search strategies presented by Mourão et al. (2020). However, as the research progressed, several other findings emerged as a result of the overall objective. Thus, the contributions of the article are summarised as follows:

- Two new concepts are introduced to mitigate the different judgements of reviewers in an SLR. The first concept is so-called *wild cards* to allow for taking an article to full text assessment even if it does not meet the formal inclusion criteria used to filter the studies to move to the next stage. Secondly, we introduce the concept of *borderline articles*, which are articles that were close to being included in the SLR. The objective of keeping track of them is to allow others to make their own judgement concerning these articles.

- A replication of an SLR on industry–academia collaboration is presented. The replication includes approximately 30% more relevant articles than the original SLR, and hence it provides an added value in relation to the original SLR on industry–academia collaboration (IAC).

- A comparison between the original SLR and the replication is provided to discuss the differences and similarities.

- The replication, using a hybrid search strategy, identified substantially more articles than the original SLR. It illustrates the value of the hybrid search strategy with a systematic database search to create a start set and then to use snowballing. Thus, a hybrid search strategy is a competitive alternative when conducting systematic literature reviews.

- Based on the outcome, in this particular case, a new hybrid search strategy is put forward for further research. We have chosen to call it an adaptive hybrid search strategy.

Here, the hybrid search strategy is implemented using a search string in Scopus to identify article candidates for inclusion. These articles are assessed, and for those included, both backward and forward snowballing are conducted using the guidelines for snowballing Wohlin (2014). In addition, to evaluate more cost-efficient hybrid methods, three alternatives of snowballing are considered. Mourão et al. (2020) introduced the four alternatives. They are summarised in Section 3.2.

The main advantage of a hybrid search strategy is that it is intended to mitigate some of the drawbacks of using different implementations of databases or indexing services and ensures that we have a good start set for snowballing.

The remainder of this article is organised as follows. In Section 2, we introduce related work, and the research design is presented in Section 3. Section 4 details the preparations for the evaluation, followed by the conduct of the replication in Section 5. Next, Section 6 presents the results in the form of an evaluation, i.e., a comparison between the original SLR and the replication, and threats to validity are discussed in Section 7. Finally, Section 8 concludes the work concerning the evaluation of the hybrid search strategy and provides suggestions on future work.

## 2. Related work

This section chronologically summarises the state–of–the–art of search strategies for SLRs in Software Engineering (SE). As demonstrated in Figure 1 and Table 1, up to now, database search (DBS) and snowballing (SB) are the leading approaches adopted in SE to search for primary studies for SLRs.

Dieste et al. (2009) analysed the effects of using different terms and combinations of terms to find an optimum search strategy for use in SLRs of SE experiments. In total, 29 search strategies were investigated, and they concluded that optimising search strings for retrieving relevant SE experiments is not a straightforward task.

Skoglund and Runeson (2009) proposed and evaluated a search strategy that uses semantic information in references between articles to find relevant articles. The strategy is composed of four steps: (1) identification of a "take-off article" – a relevant article on the SLR topic, which is systematically chosen; (2) candidate articles referenced by the "take-off article" – the reference list of the "take-off article" is analysed to reveal other relevant articles and the reference list of each revealed article is also analysed; (3) identification of "cardinal articles" – those articles referenced more than others; and (4) articles from external sources referencing the "cardinal articles" – these external articles may be identified using digital libraries. They evaluated their strategy over three SLRs and observed that the results were not satisfactory for two of the SLRs in terms of precision and relative recall.

Table 1: Chronological view of search strategies for SLRs in SE.

| Comparison | Hybrid search? | Reference |
|---|---|---|
| Searches exercising terms | | Dieste et al. (2009) |
| Reference-based search | | Skoglund and Runeson (2009) |
| MS with DBS | | Kitchenham et al. (2010) |
| MS with DBS | | Zhang et al. (2011) |
| DBS with Google Scholar + BS | | Jalali and Wohlin (2012) |
| Google Scholar + BS*FS with DBS | | Wohlin (2014) |
| Google Scholar + BS*FS with DBS | | Badampudi et al. (2015) |
| FS with DBS | | Felizardo et al. (2016) |
| FS with DBS | | Wohlin (2016) |
| DBS with Scopus + BS; Scopus + FS; Scopus + BS//FS | C | Mourão et al. (2017) |
| FS with DBS | | Felizardo et al. (2018) |
| FS with DBS | | Mendes et al. (2019) |
| DBS with Scopus + BS*FS; Scopus + BS//FS; Scopus + BS+FS; Scopus + FS+BS | C | Mourão et al. (2020) |
| Legend: Manual Search (MS); Database Search (DBS) and Backward/Forward Snowballing (BS/FS) | | |

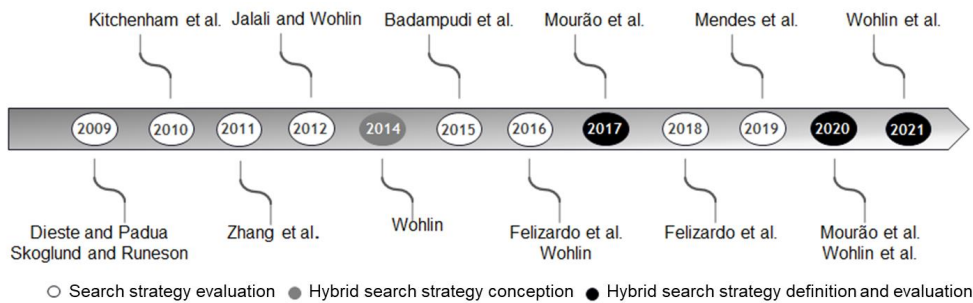

Figure 1: Search strategies for SLRs in SE shown in chronological order.

Kitchenham et al. (2010) undertook two participant-observer case studies aiming to compare both the use of manual searches with broader searches and the quality of the articles retrieved. The original SLR replicated in their study is about SLRs in SE, and the manual search covered selected journals and conferences. In the replicated study, searches including digital libraries and general indexing services were carried out. Their results indicated that searches in digital libraries and indexing services find more articles than manual searches. Still, broader searches may require more time and effort, and the articles found might be of low quality, as highlighted by Kitchenham et al. (2010).

Table 2: Main results of search strategies comparison.

| Comparison | Main result | Reference |
|---|---|---|
| Different terms | It is not straightforward to optimise a search strategy for retrieving SE experiments articles. | Dieste et al. (2009). |
| Reference list | The reference-based strategy seems to work better in specific SE topics. | Skoglund and Runeson (2009) |
| MS and DBS | DBS searches were able to find more articles than MS, but potentially with lower quality. | Kitchenham et al. (2010). |
| MS + DBS | A QGS-search strategy. | Zhang et al. (2011) |
| DBS and BS | Similar results for both search strategies. | Jalali and Wohlin (2012). |
| SB and DBS | SB may be a potential alternative to DB searches. | Badampudi et al. (2015). |
|  | Efficiencies of SB and DB searches are equivalent. | Wohlin (2014) |
| FS and DBS | DBS and FS are comparable in finding articles. | Felizardo et al. (2016, 2018); Mendes et al. (2019) and Wohlin (2016) |
| DBS with hybrid search strategies | Hybrid strategies may be a contender for SLRs. | Mourão et al. (2017) and Mourão et al. (2020) |

Manual Search (MS); Database Search (DBS); Quasi-Gold Standard (QGS) Snowballing (SB) = Backward/Forward Snowballing (BS/FS)

Zhang et al. (2011) proposed a search strategy for retrieving articles for SLRs in SE. The strategy is based on the concept of *"quasi-gold standard"* (QGS) and integrates manual and database searches. Initially, a pool of articles is manually identified. Next, this set of articles is used to elicit relevant terms to elaborate a search string for database search and validate the search performance (quasi-sensitivity). The performance is calculated by the number of relevant articles retrieved from the manual search through the database search, divided by the pool size of QGS. On the one hand, the search process terminates if the quasi-sensitivity is ≥ 80% (acceptable performance), and the results from the database search and the QGS are merged. On the other hand, the search string is refined until the performance becomes acceptable. Two participant-observer case studies were conducted to demonstrate and evaluate the QGS–search strategy, which was considered to improve the rigour of search processes in SLRs.

Jalali and Wohlin (2012) compared two SLRs on Agile practices in GSE using DBS and Backward Snowballing (BS), respectively. The same authors performed the SLRs for the same time interval to enable comparison. The start set of articles for the BS approach was generated through a search in Google Scholar, and then BS was applied based on the articles found. The search terms and keywords were kept as similar as possible in both studies. They noted that regardless of strategy, most included articles were the same. Furthermore, the conclusions found in both studies were quite similar, concluding that the SLR results were not dependent on the search strategy. They also learned that SB might be more efficient when the keywords for searching include general terms since it reduces the amount of noise in DBS.

Wohlin (2014) established guidelines for conducting snowballing (SB) as a search strategy for SLRs. The strategy defines running backward snowballing and Forward Snowballing (FS) in iterations. The application of SB was also compared with an SLR that originally used DBS as a search strategy. The results indicated that SB could replace the search in several different databases, and hence being an alternative search strategy to use when conducting SLRs. Furthermore, Wohlin (2014) puts forward the idea of a hybrid search strategy for further research, see Figure 1.

The study of Badampudi et al. (2015) complements the studies by Jalali and Wohlin (2012) and Wohlin (2014). It evaluated the effectiveness (number of articles included in relation to the total number of articles reviewed) and the reliability (the ability to capture all relevant articles) of using SB (BS and FS) as a search strategy in SLRs when compared with DBS. In their study, DBS and SB were conducted independently by different researchers in the same time period. Like Jalali and Wohlin (2012), the SB search was carried out by first creating a start set using Google Scholar. Next, BS and FS were applied based on the articles found in an iterative manner. The outcomes of the study by Badampudi et al. (2015) are similar to those reported by Jalali and Wohlin (2012) and Wohlin (2014); they found that the effectiveness of SB is comparable to DBS and that its reliability is highly dependent on the creation of a suitable start set.

Between 2016 and 2020, Felizardo et al. (2016), Wohlin (2016) and Mendes et al. (2019) have investigated SB as a search strategy for updating SLRs. They used FS to replicate a second-generation study conducted using double DB searches; i.e., DB searches were performed twice covering different time periods, i.e., for the original SLR and its update. The key conclusion is that FS can find relevant articles in updating SLRs in SE. Felizardo et al. (2018) evaluated the use of specific and generic (e.g., Scopus or Google Scholar) databases/services for applying FS to update SLRs. They concluded that using an indexing service (Google Scholar) is sufficient to find articles. These studies together enabled the definition of guidelines focused on recommendations for a search strategy specific for updating SLRs Wohlin et al. (2020). The main recommendations include employing FS using Google Scholar and using the SLR and its primary studies to compose the starting set of articles for the FS.

As shown in Table 2, only two studies, Mourão et al. (2017) and Mourão et al. (2020), have addressed the use of hybrid search strategies in SLRs in the SE area. In 2017, Mourão et al. (2017) proposed and evaluated one hybrid search strategy. More recently, in 2020, the research was extended to four hybrid search strategies that combine DBS and SB in different ways. These four strategies are described in detail in Section 3. The authors compared the outcome of DB searches, snowballing, and hybrid strategies and concluded that using a hybrid search strategy involving a representative digital library (e.g. Scopus) and parallel or sequential snowballing may be an appropriate alternative for searching for candidate articles in SLRs.

Differently from the studies mentioned above, which in most cases directly contrasted DBS with SB, our study evaluates hybrid strategies. The research reported in this article is aligned with the two previously mentioned studies by Mourão et al. (2017) and Mourão et al. (2020). However, in particular, we evaluate the four hybrid search strategies presented by Mourão et al. (2020) over one existing SLR in SE. The SLR adopted a database search strategy, followed by some complementary searches using BS and FS, and further complementary searches in some specific venues for expanding the candidate articles. Finally, we also suggest a new hybrid search strategy for further investigation.

## 3. Research design

The research team for evaluating the hybrid search strategies has previously carried out research together on SLRs. It includes conducting an update of an existing SLR, which is closely related to replicating an SLR. Thus, the team has collective experience in this type of research.

*3.1. Research questions*

The following two research questions were posed at the start of the research:

- RQ1: How many articles do the replication using a hybrid search strategy find in relation to the original SLR?

- RQ2: How do the four alternative hybrid search strategies compare to each other concerning the number of articles identified?

*3.2. Four hybrid search strategies*

The four hybrid search strategies all start with having a start set using Scopus. Then, we have the following four alternatives for hybrid search strategies, as described in Mourão et al. (2020):

1. **Scopus + BS*FS**: New articles are identified through backward and forward snowballing based on the start set. This is done for all articles meeting the inclusion criteria. This alternative is the full-fledged search strategy according to the guidelines for snowballing in Wohlin (2014).
2. **Scopus + BS /FS**: In this alternative, backward and forward snowballing run as two separate processes using the same start set. In other words, the articles obtained by backward snowballing are not subject to forward snowballing and vice-versa. This strategy was first introduced by Mourão et al. (2017) to increase the precision without compromising recall.
3. **Scopus + BS+FS**: Here, backward snowballing is conducted using the start set following the guidelines for snowballing in Wohlin (2014). After finishing all backward snowballing iterations, forward snowballing begins by using the start set from Scopus and the articles found through the backward snowballing. Backward snowballing is not done for the articles found through forward snowballing.

4. **Scopus + FS+BS**: The fourth alternative is similar to the previous option, although starting with forward snowballing. Thus, forward snowballing is conducted using the start set following the guidelines for snowballing in Wohlin (2014). After finishing all forward snowballing iterations, backward snowballing begins by using the start set from Scopus and the articles found through the forward snowballing. Forward snowballing is not done for the articles found through backward snowballing.

The objective is to use the first hybrid search strategy and then carefully keep track of when different articles are found to "simulate" the other three hybrid search strategies.

*3.3. Research approach*

To address the research questions, we defined a research approach inspired by the general process for problem–solving, as formulated by Agnew and Pyke (2007), which has three steps: observe–think–test. The three steps are reformulated herein as prepare–conduct–evaluate to align with our research focus. Thus, the research approach is as follows:

**Prepare**: A search was conducted in Google Scholar to identify candidate SLRs, using the following search string: "systematic literature review software engineering". The search used the time interval 2015-2019, and patents and citations were not ticked. The criteria and process for selecting an SLR for replication are presented in Appendix A. It resulted in selecting an SLR investigating industry–academia collaboration in software engineering, authored by Garousi et al. (2016).

Once the SLR was decided upon, it was distributed among the authors. It was agreed that we should only look at the sections preceding the presentation of the results. Reading part of the original SLR was needed to understand the research design and replicate the selection process and, in particular, the inclusion and exclusion criteria. The first author prepared Excel sheets to support the evaluation, which mitigated some of the challenges with the authors being distributed.

**Conduct**: A start set for snowballing (both backward and forward snowballing) was identified for the topic of the selected SLR. The start set was determined by searching in Scopus and evaluating the articles found in the search. Scopus was chosen motivated by the results found in Mourão et al. (2020). The inclusion of articles in our replicated SLR was based on the criteria documented in the published original SLR. Once a start set was identified, backward and forward snowballing were conducted to determine further articles to include using the guidelines by Wohlin (2014).

**Evaluate**: An analysis was carried out to address the research questions. The comparison was made concerning the included articles and the coverage of the topic of the SLR. Based on the analysis, the observations from the evaluation are reported, and the different search strategies are contrasted.

The three steps in the evaluation process are described in more detail in the following three sections.

## 4. Prepare

*4.1. Essential aspects of the selected SLR*

The objective of the selected SLR by Garousi et al. (2016) is stated as follows: "To identify (a) the challenges to avoid risks to the collaboration by being aware of the challenges, (b) the best practices to provide an inventory of practices (patterns) allowing for an informed choice of practices to use when planning and conducting collaborative projects." Garousi et al. (2016). Thus, articles describing collaboration per se between academia and industry should not be included since the focus is on challenges and best practices in the collaboration. Furthermore, the authors of the selected SLR described the context of their SLR as being experiences and lessons learnt in industry–academia collaboration, as reported either by researchers or practitioners. The focus of the chosen SLR is summarised in three research questions by the authors. They look at collaboration models between industry and academia, challenges and impediments that have been highlighted, and finally, patterns in terms of best practices. To identify articles, the authors used the following search process:

- The authors searched in three different sources: IEEE Xplore, ACM Digital Library and Google Scholar.

- Fifty-four (3*2*3*3) search strings were used by combing four different listings and searching for all combinations. The listings can be found in Table 3.

- • The searches were conducted in January and February 2015. Only articles, available at this time, were included in the pool of potential studies for inclusion in the SLR.

- To complement the searches in IEEE Xplore, ACM Digital Library and Google Scholar, and to ensure that they did not miss any relevant articles, the authors randomly selected five articles from the searches and conducted backward and forward snowballing (Wohlin, 2014). It is not reported which five articles were included in the snowballing.

- Finally, the authors also looked at some specific venues for additional articles. The selected venues are not reported in their article.

| Listing 1 | Listing 2 | Listing 3 | Listing 4 |
|---|---|---|---|
| Industry | Academia | Collaboration | Software engineering |
| Practice | Theory | Relationship | Software |
| University | | Relation | IT |

Table 3: List of keywords for creating search strings.

The search strategy is essential since our hybrid approach should be compared to the strategy applied by the original SLR's authors.

When it comes to the inclusion/exclusion criteria for articles, the following is stated in Garousi et al. (2016):

- The main criterion concerns whether a given study presents relevant findings for industry and academia collaboration in software engineering.

- Only articles written in English are included.

- Only articles available electronically are included.

- If a conference article has a more recent journal version, then the journal article is included, and the conference article is excluded.

- Only the most recent article is included if multiple studies with the same title by the same authors are found.

The inclusion/exclusion process is described as follows:

- All three authors looked at the pool of potential articles to be included. It is unclear how the initial pool was identified based on the search process, i.e., the initial screening of articles is not described.

- For each article, the authors assigned individual scores: 0 – exclude, 1 – uncertain, and 2 – include.

- The first scores were based on reviewing the title, abstract and keywords of each article. If this was insufficient to assign a score, the authors looked at the full article. However, it is unclear when they looked at the full articles, i.e., it is not stated if each individual decides it or whether it is a joint decision.

- Articles were tentatively included as candidates if the total score was four or higher and excluded if the total score was lower than four. Final inclusion/exclusion was based on looking at the full articles, which is needed independently to assess the contribution of articles in relation to the research questions in a regular SLR.

*4.2. Our preparations*

*4.2.1. Introducing two novel concepts*

Two novel concepts were introduced in the reviewing process to address different judgements of the researchers assessing the articles. The two new concepts are:

- Wild cards – A *wild card*, in our context, is an article that fails to qualify in the usual way, i.e., by fulfilling the score needed to be included. The concept is taken from sports, where certain persons may be invited to participate in a tournament even if not meeting the qualification criteria. Each reviewer is allowed to nominate a wild card as described in Section 4.2.4.

- Borderline articles – A *borderline article* is defined as an article receiving a score of three in the full text assessment. To achieve a score of three, either all three reviewers are uncertain, or one reviewer wants to include the article, and one reviewer is uncertain. Thus, the articles being assessed are placed in three categories: include, exclude, and exclude, but borderline. Articles in the third category are kept track of separately. The objective is to provide transparency into which articles were close to being included. Furthermore, it allows readers of our review to assess the borderline articles themselves and decide whether they are relevant to them or not.

We believe that both concepts embrace differences in opinions instead of, for example, having two persons convincing a third person that a paper should be excluded.

*4.2.2. Supporting Excel sheets*

The first author developed an Excel sheet to support the identification of the start set. The Excel sheet included links to the articles identified in Scopus to help the individuals make their judgements concerning these articles. Furthermore, Excel sheets were developed for both backward and forward snowballing to support the reviewers.

*4.2.3. Reviewers of articles*

The first three authors of the article acted as reviewers of the articles found through the search in Scopus and also took part in the backward and forward snowballing. It was essential to involve three researchers to mimic the process used by the authors of the original SLR. The fourth author acted as a backup in the reviewing process.

*4.2.4. Inclusion/exclusion process*

All articles found were assigned scores, as done in the original SLR, i.e. first on title, abstract and keywords, and then for tentative candidates, the full article was assessed. As above-mentioned, three researchers assessed the articles identified. Note that if a reviewer perceives that it is infeasible to give a score without looking at the full article, the score assigned is one for "uncertain".

Moreover, the inclusion/exclusion process in the original SLR was adapted as follows. Each reviewer was allowed to nominate a "wild card" if it was perceived that the full article should be assessed even if it did not meet the score for an article to be included for full text assessment.

If an article was nominated as a wild card by more than one reviewer, the reviewers having nominated the same article were allowed to nominate an additional article each. It was done until the number of included wild card articles was at least the same as the number of reviewers.

Articles selected for full text assessment were reviewed by the same reviewers as for the initial screening. Furthermore, the articles reviewed in full text were scored similarly as in the screening of articles on title, keywords and abstract. No additional wild cards were used after full text review, i.e., only articles meeting the scoring threshold were included as a primary study in the SLR. However, articles receiving a score of three were kept track of separately in a list of "borderline articles".

## 5. Conduct

*5.1. Search in Scopus*

In the hybrid search strategy, the first step is to identify a start set for snowballing. As stated in the research design, it was decided to use Scopus to search for articles to form the start set.

The search in Scopus was done for the five years preceding the date when the selected SLR was conducted, i.e., not published. It was decided to have a start set of at least five articles, and preferably ten articles, to start the snowballing procedure. Having only a few articles may bias the snowballing and result in missing relevant articles. Thus, a sufficiently large subset is needed. However, it is hard to determine an optimal number since it depends on the number of articles in the area (which is unknown upfront) and the number of active researchers in the area (since authors tend to at least cite their own relevant articles). The start set should be articles that are to be included among the primary studies in the SLR. If less than five articles were found, the objective was to increase the time interval by one year at a time until the start set had at least five articles.

The candidate articles for the start set were found using the following process:

1. Given the area of the selected SLR to replicate, the following search string was formulated to be used in Scopus: industry AND academia AND collaboration AND software AND engineering.
2. Given that the selected SLR was conducted in early 2015, although published in 2016, the first search is done for the time interval 2010-2014.
3. Only articles published in either journals or conferences are included, which also means that review articles published in either journals or conferences are included. Articles of types: book, book chapter, conference review and notes are excluded from the search in Scopus.

The total number of articles found in Scopus was 40 articles. Their titles were put into the Excel sheet with links to the articles, based upon the listing generated by Scopus. It ensured that the authors used the same information when providing scores for the different articles.

*5.2. Identification of start set*

The articles found in the search were evaluated based on the inclusion and exclusion criteria in the original SLR and rated as described in Section 4.2.4. Given that snowballing should only be conducted on articles to be included in the final set of articles, it is necessary to look at the full articles identified in the scoring procedure, including the wild cards, before performing the next step in snowballing. However, it was unclear from the original SLR whether or not articles relating to the collaboration between industry and academia in education should be included or excluded. To obtain a clarification, one of the authors of the original SLR was contacted. We were informed that the focus was on research collaborations. Thus, given that the objective was replication, we also focused on articles concerning collaboration in research.

In total, 15 articles fulfilled our inclusion criteria to go into full text reading. It included twelve articles being selected based on the scores and three articles as wild cards. The 15 articles were assessed based on scores by the three reviewers, i.e., no wild cards were used when doing full text reading.

It resulted in nine articles being included. All nine articles included had a total score of six (each reviewer gave a score of two, see Section 4.1), which means that all three reviewers wanted to include the articles. Thus, there was a consensus for inclusion among the three reviewers. Six articles were excluded. The opinions of these six articles varied. For five of the articles, one reviewer wanted to include the article. However, it should be noted that only one article had a score of three, and hence only one article was borderline to be included. The articles are listed separately as described above. It is also worth mentioning that none of the three wild cards nominated was included. Thus, the assessment on title, abstract and keywords is well aligned with the assessment on full text reading.

The different views among the three reviewers were not critical for the main objective in the article, i.e., to assess and compare the various search strategies applied in the original SLR and the replication presented here. Thus, even if an article was excluded, it is documented that it was found. This will be considered when comparing the search strategies used in the original SLR and the replication.

Although investigating only journal and conference articles when identifying the start set in Scopus, it was also decided to include book chapters and workshop articles when conducting snowballing. We did not ask the authors of the original SLR since we wanted to be as independent as possible when conducting the search. We made an exception when asking concerning industry–academia collaboration in education, where the answer was a simple yes/no. If asking for publication types included, we may get into a discussion concerning publication types, which we wanted to avoid. Furthermore, it is substantially easier to remove them later than add them. Book chapters and workshop articles were included under the assumption that they were peer–reviewed. We did not consider other types of publications such as, for example, books and theses, since they have most likely not been through the same peer–review as research articles, including book chapters.

*5.3. Backward snowballing*

In backward snowballing, the following process was used:

1. An Excel sheet was constructed with a listing of the articles included based on the search in Scopus. These articles formed the start set for the first round of backward snowballing. In the second round, articles included from the first round of backward and forward snowballing are included, and so forth.
2. Each reviewer was asked to go through the reference lists of the included articles. Articles that are judged to definitively be out of the scope of the systematic literature review are not moved to the Excel sheet. The judgement is done based on the article's title and how and where it was referred to in the article. Articles that potentially could be of interest are put into a personal Excel sheet by each reviewer.
3. Each reviewer assesses the articles put into their respective Excel sheet based on title, abstract and keywords using the same procedure as when reviewing the articles found through the search in Scopus, as described in Section 4.2.4.
4. The assessment based on the title, abstract and keywords is coordinated. It means that if one of the reviewers gave a score of 1 or 2 for an article not assessed by one or two of the other reviewers, it also has to be assessed by those who have not assessed it. Thus, all articles with scores of 1 or 2 from at least one reviewer were also assessed by all reviewers.
5. The full text assessment was conducted as described in Section 4.2.4.

It should be observed that we were careful to track all included articles in terms of when they were found. It is evident that there is a need to keep track of when an article was identified for inclusion for the first time. However, to evaluate all four hybrid search strategies, we need to keep track of all the instances an article is found.

*5.4. Forward snowballing*

In forward snowballing, the following process was used:

1. The title of each article was put into Google Scholar to find the article. For each title, the citations were identified also using Google Scholar. The search for citations was limited to articles published in 2014 or earlier. Patents and citations were unticked.
2. Links to the peer–reviewed articles citing each article were put into an Excel sheet by the first author to simplify the assessments of the citing articles. It means that non–peer–reviewed publications, for example, books and theses, are removed before adding the publications into the Excel sheet.
3. The articles were assessed as described in Section 4.2.4.

*5.5. Articles identified*

The replication identified a total of 43 articles published in 2014 or earlier, which should be contrasted with the 33 publications listed in the original SLR. Figure 2 illustrates the articles identified based on the nine articles of the start set throughout the snowballing iterations. Furthermore, ten articles were listed as borderline articles. The articles included in the start set and after each backward and forward snowballing iteration are listed in Appendix B, which also lists the borderline articles. The results are further elaborated in Section 6.

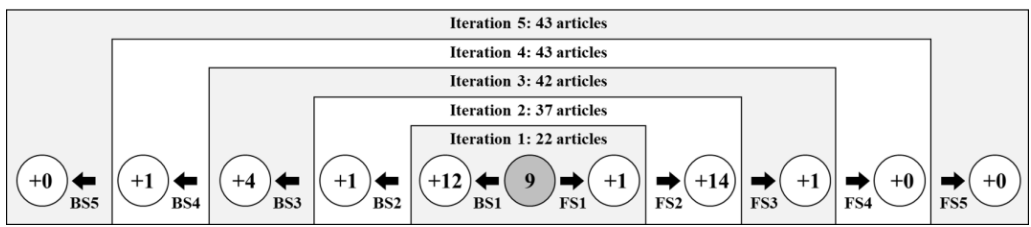

Figure 2: Articles identified throughout the snowballing iterations based on the start set.

## 6. Evaluate

The results include a discussion on the two new concepts introduced in Section 6.1. Some reflections concerning conducting the evaluation through a comparison with the original SLR are provided in Section 6.2. In Section 6.3, a comparison of the original SLR and its replication are presented in terms of the articles identified through the two different search strategies. Finally, in Section 6.4, the different hybrid search strategies are compared.

### 6.1. The two new concepts

Both new concepts, i.e., wild card and borderline article, were found helpful. One of the wild card articles is among the included articles after the full text assessment. The article by Punter and van de Laar (2010) was nominated as a wild card and then included after full text assessment. This article was not included in the original SLR. Furthermore, four nominated wild cards made it into being borderline articles. The four articles have numbers 70, 73, 115, and 140 in Appendix B. Article 140 is included in the original SLR, while the other three articles are not.

The wild card concept resulted in one article, which otherwise would have been excluded, to be included. Furthermore, it helped identify four more borderline articles. The concept of borderline articles is also helpful since it illustrates to readers which articles were close to being included. In this way, readers may decide for themselves whether or not the articles contain useful findings. In summary, we argue that both concepts are valuable additions to the current standard way of conducting SLRs. Wild cards help mitigate different opinions among the reviewers, and borderline articles provide additional information to the readers of an SLR.

### 6.2. Reflections on evaluation

Before comparing the outcome of the original SLR and the replication in more detail (or any two SLRs on the same topic), four aspects need to be separated:

1. Inclusion and exclusion criteria: Although these are stated in the original SLR, they may be interpreted differently by different research teams. Implicit agreements between the researchers involved may not be fully captured in the criteria documented, which is a challenge when comparing two SLRs or replicating an SLR.

2. Search strategy: The search may have been done in different ways, as here, when the search strategy was deliberately different, i.e., to compare the search strategies.
3. Inclusiveness: This refers to what is needed to take an article to the next assessment level. We distinguish at least three levels here. First, articles may be assessed only on the title, i.e., the immediate information available when searching in, for example, Google Scholar, or when looking at articles in a reference list. The second level is concerned with assessing, for example, primarily the abstract, but in conjunction with the title and keywords. The third level means that the full text for the articles is assessed.
4. Judgement: The final judgement concerning inclusion or exclusion is done based on full text assessment. Different researchers may not judge an article similarly since it is a subjective task to assess an article.

The execution of the approach in the four items mentioned above differed between the original systematic literature review and the replication presented here. In some cases, the differences were intentional and in other cases unintentional.

Given that it was unclear whether industry–academia collaboration related to education should be included or excluded, we checked with the authors before conducting the replication. We were informed that the focus was on research collaboration, and hence articles concerning collaboration in education were excluded. Another aspect, which we learnt after having conducted the replication, was that in the original SLR, only articles having their primary focus on industry–academia collaboration in software engineering were included. Thus, articles describing a collaborative research effort in different areas of software engineering, such as requirements engineering, design or testing, and including, for example, a section on reflections concerning the collaboration were, according to one of the authors of the original SLR, excluded in the original SLR. Given that the main inclusion criteria in the original SLR were formulated as follows: "Does a given study present findings relevant for IAC in SE?" we included articles not solely focusing on industry–academia collaboration in software engineering. We have separated the two types of articles in the analysis to allow for a fairer comparison between the two SLRs (original and replication). We report both types of articles for two reasons. First, we perceive that the original SLR does include some articles not solely focusing on industry–academia collaboration.

Secondly, we judge that some of the articles not exclusively concentrating on collaboration contain essential lessons learnt.

Second, the search strategy was intentionally different and part of the research questions. Thus, this is the aspect we want to compare, although it is affected by the other three items above listed.

Concerning the third item, after exchanging emails with the first author of the original SLR, it is clear that we have taken a more inclusive approach in the replication to not miss any article due to its title and abstract not being sufficiently informative. To a large extent, our inclusiveness is a consequence of not only including articles with their primary focus on industry–academia collaboration. Thus, we have also included articles with general experiences concerning industry–academia collaboration, although it is not the article's primary focus. Focusing solely on articles with the primary focus on industry–academia collaboration makes the process more efficient, but increases the risk of missing some essential experiences. This illustrates the influence of how the scope of an SLR is formulated. Although not sufficiently clearly expressed in the article, the more focused scoping of the original SLR explains some of the differences in the number of articles assessed by the different research teams. Whether higher inclusiveness is worth the extra effort is a subject for further research.

Finally, the judgement is likely to be different due to being done by humans and hence being subjective. Our two new concepts, i.e., wild cards and borderline articles, are intended to help mitigate the issue and make possible differences transparent. Moreover, to help understand the potential differences between the two teams (original SLR and replication), we have been provided access to articles assessed but being excluded in the original SLR.

*6.3. Comparison of the original SLR and its replication*

Table 4 presents a comparison of the publications listed in the original SLR and the replication. The first row (below the row with the headings) in the table highlights the number of publications included by each research team. The percentages show that the replication found 30% more relevant publications than the original SLR concerning industry–academia collaboration in software engineering. It is worth mentioning that the number of publications included by both the original SLR and its replication is 20 articles.

On the second row, four publications listed in the original SLR have been removed since they are not judged to have been peer-reviewed. It includes

Table 4: Comparison original and replicated SLR.

| Comparison | Original SLR | Replicated SLR | Percentage |
|---|---|---|---|
| Listed in respective SLR | 33 | 43 | 30% |
| Remove non-peer-reviewed publications | 29 | 43 | 48% |
| Remove articles excluded in replication | 24 | 43 | 79% |
| Remove after assessment articles included in original and not in replication | 22 | 43 | 95% |
| Remove articles not having a primary focus on IAC | 19 | 36 | 89% |

two keynotes, one book and a contribution to a column in IEEE Software.

Concerning this comparison, the original SLR includes nine publications not included in the replication, i.e., given that 20 articles were found by both the original SLR and the replication. However, five of these have been assessed when conducting the replication. Two of them are included among the borderline articles, i.e., Baldassarre et al. (2013) and Morris et al. (1998). The third article, by Connor et al. (2009), focuses on industry–academia collaboration in education. Hence it should be excluded based on the information from one of the authors of the original SLR. The fourth article by Rombach et al. (2008) was excluded at the abstract level. It received a score of zero from the reviewers conducting the assessment for the replication. Finally, the fifth article by Lamprecht and van Rooyen (2012) was excluded after full text reading with a score of two. The latter article addresses IPR concerns when industry and academia collaborate, and it is in the context of regulations in South Africa. Thus, the article does not address the actual collaboration. Moreover, it is focused on one aspect related to the relationship between industry and academia, and it is in the context of regulations in one country, and hence it was excluded. The removal of these five articles from the original SLR led to the outcome presented in the third row in Table 4.

The remaining four articles included in the original SLR and not found in the replication were assessed by the research team of the replication. It resulted in the following, an article by Franch et al. (2012) was excluded with a score of one in the full text reading, and an article by Osterweil et al. (2008) was excluded with a score of zero on the abstract level given that it discusses the impact of research on practice and not industry–academia collaboration. Thus, given that these two articles were not found relevant, we obtain the results on the fourth row in the table. The other two articles were assessed as being appropriate for inclusion. It is the articles by Raschke et al. (2014) (with a score of four) and Krishnan et al. (2009) (with a score of five).

Finally, given that we were informed by one of the original SLR's authors that only articles having the main focus on industry–academia collaboration in software engineering were included, we looked at all articles included by both the original SLR and the replication. It resulted in removing three articles in the original SLR and seven articles in the replication. The results from this removal can be found in the fifth row of Table 4. However, we believe that some of the articles not primarily focused on industry–academia collaboration, but including essential experiences (typically in a separate section of the article) from collaboration, should be included. Based on this, we find the comparison on the fourth row most relevant. It includes 22 peer-reviewed articles in the original SLR and 43 peer-reviewed articles in the replication. Given the overlap of 20 articles, the "superset" includes in total 45 peer-reviewed articles, including essential findings concerning industry–academia collaboration published in 2014 or earlier. In summary, the targeted findings, as described in the original SLR, include collaboration models, challenges and best practices.

Conducting SLRs comes with substantial effort to assess articles independent of the search strategy. On the one hand, we have one search in Scopus, while the authors of the original SLR have 54 different search strings to run in two databases (IEEE and ACM) and one indexing service (Google Scholar). On the other hand, we have assessed all references and citations when conducting snowballing in the hybrid search strategy. It is unfortunately impossible to compare the effort, and hence also the cost–efficiency. However, if trying to find *all* relevant literature (in English), then effectiveness ought to be prioritised over efficiency. From an effectiveness perspective, the hybrid search strategy outperforms the database search and the complementary searches conducted in the original SLR, independent of how the comparison is made, see Table 4.

*6.4. Comparison of the hybrid search strategies*

When comparing the four alternative hybrid search strategies (see Section 3.2), it becomes evident that their performance is highly dependent on the start set. As presented above, we found 43 articles when using the first alternative strategy. The other three alternatives try to make the work more efficient by not conducting all forward and backward snowballing combinations. The results concerning the number of articles identified by the four alternative search strategies are shown in Table 5.

Table 5: Comparison of hybrid search strategies.

| Search strategy | Number of articles |
|---|---|
| Strategy 1: Scopus + BS*FS | 43 |
| Strategy 2: Scopus + BS\|\|FS | 23 |
| Strategy 3: Scopus + BS+FS | 38 |
| Strategy 4: Scopus + FS+BS | 23 |

Based on the outcome presented in Table 5, we made the following observation. In this case, most articles are found in the first round of backward snowballing and the second round of forward snowballing. Thus, running backward and forward snowballing in parallel means that they do not benefit from each other. When having a round with few articles in backward or forward snowballing, we risk that the procedure stops early. Given that many articles, in our case, are found with backward snowballing in the first round, the best option is to use search strategy 3 since the articles found in backward snowballing will be used when conducting forward snowballing. Thus, search strategy 3 is superior, in our case, if trying to make the search more efficient.

However, as indicated above, the performance of the different hybrid search strategies is dependent on the start set and publication patterns over the years in the investigated time interval. In our case, we created a start set by searching in Scopus in the time interval 2010-2014. Overall, we are interested in articles published in 2014 and earlier. Therefore, the search in Scopus is focused on relatively new articles published in 2014 or earlier. Thus, it is no surprise that backward snowballing is superior in the first round. Then, as we identify older articles, forward snowballing performs well since it looks at newer articles relative to those found in backward snowballing. It indicates that alternating between backward and forward snowballing may be an option (in this case). If starting with one round of backward snowballing and then continue with one round of forward snowballing, and continue alternating between the two ways of conducting snowballing, we identify 40 articles. It does not mean that we find all 43 articles, but it finds more articles than the hybrid search strategies 2–4.

A potential way forward is to have an adaptive hybrid search strategy depending on the outcome. In this case, the adaptation would be to have an alternating hybrid search strategy since we started with relatively few articles in the start set, and hence going backwards first is most likely the best option. We may also consider the number of references in relation to the number of citations to the articles in the start set to decide how to make the searches more efficient. However, this is an area for future research.

## 7. Threats to validity

In this case, the threats to conclusion validity are primarily related to selection bias and evaluator bias. The selection of a specific SLR may bias towards the hybrid search strategy. However, the SLR was selected using a set of criteria directed towards the content of the SLR and did not favour a hybrid search strategy. Furthermore, the authors of the selected SLR have been very helpful with information concerning the original SLR, which have been valuable to ensure that our interpretation of the selected SLR is as good as possible.

Furthermore, there is a risk that the individual researchers become biased, given that there is a vested interest in the hybrid search strategy. However, having three researchers conducting independent assessments on all articles using the new concept of wild cards helped mitigate individual evaluator bias. Moreover, the new concept of borderline articles makes delimitation between inclusion and exclusion more transparent to readers. Such transparency is essential to allow readers to assess the potential evaluator bias. Overall, it is judged that the design of the study and the predefined criteria for selecting an SLR to use in the evaluation help minimise the conclusion validity threats.

Another potential threat in literature reviews is publication bias, i.e., articles with specific characteristics are more often published or more often retrieved. We did not assess whether publication bias favours a particular search strategy.

## 8. Conclusions

The overall objective was to compare and evaluate a hybrid search strategy with a search using databases and indexing services. Furthermore, we wanted to compare four different alternative approaches to conducting hybrid searches. To do so, an SLR was selected for replication. The original SLR conducted searches using search string in databases (IEEE Xplore and ACM Digital Library) and one indexing service (Google Scholar). These searches were complemented with some snowballing and looking at the proceedings from some specific venues. The search strategy in the original SLR is judged to be representative of how searches often are conducted when doing an SLR, and it met a set of predefined criteria for selecting an SLR for replication.

With a start set from Scopus and then both backward and forward snowballing from all articles meeting the inclusion criteria, the full-fledged hybrid search strategy was superior to the search strategy in the original SLR. In the replication, only research articles published in journals, at conferences and workshops, and as book chapters were included. However, even when accepting keynotes, column contributions and books in the original SLR, the hybrid search strategy found 30% more articles than in the original SLR. And it performed even better if, for example, only accepting peer-reviewed articles and removing articles assessed and excluded in the replication. In summary, the hybrid search strategy is a strong contender as a search strategy when conducting systematic literature studies.

The full-fledged hybrid search strategy is better than the alternative hybrid strategies. However, it requires more effort to assess all articles identified. When comparing the full-fledged hybrid search strategy, it became clear that the success of the alternative hybrid search strategies depends on the start set, particularly how it was identified. In our case, the start set includes only relatively new articles in the investigated time interval (articles published before 2015). Hence, it is no surprise that in the first round, backward snowballing found more articles for inclusion than forward snowballing, and then it is beneficial to run forward snowballing after having done backward snowballing. Thus, the third hybrid search strategy is the second best. If wanting to save some effort, although missing some articles, it is probably best to choose an alternative hybrid search strategy based on the characteristics of the start set. Further research into adaptive (in relation to the start set) hybrid search strategies is needed.

Two new concepts are proposed to embrace the differences in judgement when assessing articles in relation to the inclusion and exclusion criteria. We introduced the concept of wild cards to allow individual reviewers to put forward an article for full text assessment even if the other reviewers think the article should be excluded when assessing the title, abstract and keywords. One wild card made it into being included in the final set of articles. The second concept is borderline articles. We suggest that articles being close to being included are kept in a separate list to allow readers to make their own judgement concerning these articles. It is noteworthy that four out of nine borderline articles come from being nominated as wild cards.

In summary, the full-fledged hybrid search strategy identified substan-tially more articles presenting models, challenges and best practices concerning industry-academia collaboration in research than the original SLR. The results strengthen the findings in Mourão et al. (2017) and Mourão et al. (2020), where it was indicated that a hybrid search strategy might be a suitable alternative to identify primary studies. Here, we conclude that the hybrid search strategy is an excellent alternative when searching for articles to include in an SLR. Furthermore, we suggest that the hybrid search strategy is complemented with two concepts, i.e., wild cards and borderline articles.

**Acknowledgement**

First of all, we would like to express our gratitude to Prof. Vahid Garousi and Prof. Kai Petersen for providing additional details concerning their systematic literature review on industry-academia collaboration.

Furthermore, we are grateful to Dr. Frank Houdek, Dr. Teade Punter and Dr. Piërre van de Laar for helping us with articles to which we did not have access without the help of the authors.

Professor Marcos Kalinowski is funded by a research grant from the Brazilian National Council for Scientific and Technological Development (CNPq), Grant #312827/2020-2.**Appendix A. Supplement – Selecting an SLR for replication**

Appendix A describes how we selected an SLR to replicate.

**Appendix B. Supplement – Included articles and borderline articles**

Appendix B provides listings of the included articles and the borderline articles, respectively. The listings show in which step of the hybrid search each article was identified.

**References**

Agnew, N.M., Pyke, S.W., 2007. The Science Game – An Introduction to Research in the Social Sciences. 7th ed., Oxford University Press.

Badampudi, D., Wohlin, C., Petersen, K., 2015. Experiences from using snowballing and database searches in systematic literature studies, in: Proceedings International Conference on Evaluation and Assessment in Software Engineering, p. 17.

Baldassarre, M.T., Caivano, D., Visaggio, G., 2013. Empirical studies for innovation dissemination: Ten years of experience, in: Proceedings International Conference on Evaluation and Assessment in Software Engineering, pp. 144–152.

Connor, A.M., Buchan, J., Petrova, K., 2009. Bridging the research-practice gap in requirements engineering through effective teaching and peer learning, in: Proceedings International Conference on Information Technology: New Generations, pp. 678–683.

Dieste, O., Grimán, A., Juristo, N., 2009. Developing search strategies for detecting relevant experiments. Empirical Software Engineering 14, 513–539.

Felizardo, K.R., Mendes, E., Kalinowski, M., Souza, E.F., Vijaykumar, N.L., 2016. Using forward snowballing to update systematic reviews in software engineering, in: Proceedings International Symposium on Empirical Software Engineering and Measurement, pp. 1–6.

Felizardo, K.R., da Silva, A.Y.I., de Souza, E.F., Vijaykumar, N.L., Nakagawa, E.Y., 2018. Evaluating strategies for forward snowballing application to support secondary studies updates: Emergent results, in: Proceedings Brazilian Symposium on Software Engineering, pp. 184–189.

Franch, X., Ameller, D., Ayala, C.P., Cabot, J., 2012. Bridging the gap among academics and practitioners in non-functional requirements management: Some reflections and proposals for the future, in: Seyff, N., Koziolek, A. (Eds.), Modelling and Quality in Requirements Engineering: Essays Dedicated to Martin Glinz on the Occasion of His 60th Birthday. Verlagshaus Monsenstein und Vannerdat, Muenster, pp. 267–273.

Garousi, V., Petersen, K., Ozkan, B., 2016. Challenges and best practices in industry-academia collaborations in software engineering: A systematic literature review. Journal of Information and Software Technology 79, 106–127.

Jalali, S., Wohlin, C., 2012. Systematic literature studies: Database searches vs. backward snowballing, in: Proceedings 6th International Symposium on Empirical Software Engineering and Measurement (ESEM), pp. 29–38.

Kitchenham, B.A., Brereton, P., Turner, M., Niazi, M.K., Linkman, S., Pretorius, R., Budgen, D., 2010. Refining the systematic literature review process–two participant-observer case studies. Empirical Software Engineering 15, 618–653.

Kitchenham, B.A., Charters, S., 2007. Guidelines for Performing Systematic Literature Reviews in Software Engineering. Technical Report EBSE-2007-01. School of Computer Science and Mathematics, Keele University.

Krishnan, P., Ross, K., Pari-Salas, P., 2009. Industry academia collaboration: An experience report at a small university, in: Proceedings Conference on Software Engineering Education and Training, pp. 117–121.

Lamprecht, S.J., van Rooyen, G.J., 2012. Models for technology research collaboration between industry and academia in South Africa, in: Proceedings IEEE Software Engineering Colloquium (SE), pp. 11–17.

Mendes, E., Felizardo, K., Wohlin, C., Kalinowski, M., 2019. Search strategy to update systematic literature reviews in software engineering, in: Proceedings Euromicro Conference on Software Engineering and Advanced Applications, pp. 355–362.

Morris, P., Masera, M., Wilikens, M., 1998. Requirements engineering and industrial uptake, in: Proceedings International Symposium on Requirements Engineering, pp. 130–137.

Mourão, E., Kalinowski, M., Murta, L., Mendes, E., Wohlin, C., 2017. Investigating the use of a hybrid search strategy for systematic reviews, in: Proceedings International Symposium on Empirical Software Engineering and Measurement, pp. 193–198.

Mourão, E., Pimentel, J.F., Murta, L., Kalinowski, M., Mendes, E., Wohlin, C., 2020. On the performance of hybrid search strategies for systematic literature reviews in software engineering. Journal of Information and Software Technology 123, 106294.

Osterweil, L.J., Ghezzi, C., Kramer, J., Wolf, A.L., 2008. Determining the impact of software engineering research on practice. Computer 41, 39–49.

Punter, T., van de Laar, P., 2010. Industrial impact and lessons learned, in: Van de Laar, P., Punter, T. (Eds.), Views on Evolvability of Embedded Systems. Springer, Dordrecht, pp. 279–299.

Raschke, W., Zilli, M., Loinig, J., Weiss, R., Steger, C., Kreiner, C., 2014. Embedding research in the industrial field: A case of a transition to a software product line, in: Proceedings International Workshop on Long–Term Industrial Collaboration on Software Engineering, pp. 3–8.

Rombach, D., Ciolkowski, M., Jeffery, R., Laitenberger, O., McGarry, F., Shull, F., 2008. Impact of research on practice in the field of inspections, reviews and walkthroughs: Learning from successful industrial uses. SIGSOFT Software Engineering Notes 33, 26–35.

Skoglund, M., Runeson, P., 2009. Reference-based search strategies in systematic reviews, in: Proceedings International Conference on Evaluation and Assessment in Software Engineering, pp. 31–40.

Wohlin, C., 2014. Guidelines for snowballing in systematic literature studies and a replication in software engineering, in: Proceedings International Conference on Evaluation and Assessment in Software Engineering, pp. 321–330.

Wohlin, C., 2016. Second-generation systematic literature studies using snowballing, in: Proceedings International Conference on Evaluation and Assessment in Software Engineering, pp. 1–6.

Wohlin, C., Mendes, E., Felizardo, K.R., Kalinowski, M., 2020. Guidelines for the search strategy to update systematic literature reviews in software engineering. Information and Software Technology 127, 106366.

Zhang, H., Babar, M., Tell, P., 2011. Identifying relevant studies in software engineering. Information and Software Technology 53, 625–637.